# Optical inspection of manufactured nanohole arrays to bridge the lab-industry gap

A. FRANCO,[1] D. OTADUY,[2] A.I. BARREDA,[3] J.L. FERNÁNDEZ-LUNA,[1] S. MERINO,[2] F. GONZÁLEZ,[3] F. MORENO,[3,*]

[1] *Unidad de Genética, Hospital Universitario Marqués de Valdecilla, 39008 Santander, Spain*
[2] *IK4-TEKNIKER, 20600 Eibar, Spain*
[3] *Group of Optics, Dep. of Applied Physics, University of Cantabria, 39005 Santander, Spain*
*\*morenof@unican.es*

**Abstract:** Metallic nanohole arrays have shown their potential as sensing tools. Important research supported by sophisticated laboratory experiments have been recently carried out, that may help to develop practical devices to be implemented in the real life. To get this goal, the gap between industry and technology at the nanoscale level must be overcome. One of the major drawbacks is the quality inspection of the manufactured nanostructured surfaces to ensure a reliable sensing. In this paper we introduce an optical method, based on the Extraordinary Optical Transmission phenomenon, for an inspection of such surfaces at industrial level. Unlike usual techniques like SEM, our method gives at the same time, a quick map of the surface homogeneity together with its local plasmonic performance. Our results show that this method is reproducible and reliable as to give a "seal of identification" and quality guarantee of the manufactured surface as basic element of a sensing device.

## 1. Introduction

Nanotechnology has revolutionized science during last decades unveiling new phenomena. In particular, when incident light goes through a periodic array of subwavelength apertures in a thin metallic film, resonance peaks are clearly observed in the transmission spectra. At their corresponding wavelengths, the intensity of the transmitted light is effectively larger than that predicted by conventional diffraction theory [1]. This phenomenon, known as Extraordinary Optical Transmission (EOT), is due to the excitation of Surface Plasmon Polaritons (SPP's) at the metal surface [2].

Nanohole arrays in metallic films have been vastly investigated for the development of biosensors devoted to medical diagnosis and to the detection of environmental pollutants. Its resonant behavior often leads to large field enhancement which is advantageous in other applications such as Surface-Enhanced Raman Scattering (SERS) [3]. In particular, due to their high sensitivity to changes of the buffer effective refractive index, they have been used as biosensors for detecting low concentrations of nano-sized biological material, such as proteins and DNA molecules suspended in a buffer medium [4]. They have been also used for detecting larger structures, in the range of hundred-nanometers, such as intact viruses [5] and exosomes [6], enabling not only to describe their molecular profile, but retrieving them for further studies [6]. Besides, they have been used for label-free cell detection [7], even if the size of these biological material is far from the nanometric range. The current high demand of non-invasive liquid biopsies is behind the interest to use the nanohole-based sensors for detecting micron-sized biological material [8]. In particular, the early detection and analysis of rare-cells in blood enables to monitor diseases, improving the patient's quality of life. For instance, the presence of circulating tumor cells in blood is correlated to metastatic diseases [9].

The incorporation of this kind of nanostructured films, henceforth called sensing chips, in biochemical sensing devices has several advantages: it allows a minimally invasive label-free sensing, the films may be easily included in size-reduced multiplexing microfluidic devices, and it is compatible with simple optical set-ups based on the detection of light transmitted through the perpendicular direction of the films surface [10].

However, it is difficult to translate these sensor devices to the technological industry, because it needs a rigorous quality-control of the sensing chips through accurate, fast and low-cost techniques. Since the lack of homogeneity in some of the nano-features may lead to undesirable optical responses [11], their characterization and further quality-control is of paramount importance for their mass production. This is particularly important when micron-sized biological material is located at specific places on the sensing chip surface [7,12].

There are three different methods for the mass production of nanohole films: 1) replication, 2) self-assembly and 3) direct pattern projection [11], but all these methods are limited by their own fabrication nature: they produce nanostructured films with inhomogeneously distributed defects. Such defects may alter the nanohole films expected optical response, because it depends on several features related to nano-structures manufacturing: hole geometry [13-15], shape of the hole walls [16,17], hole distribution within the periodic array [18], roughness and defects in their neighborhoods [19, 20] and number of illuminated nanoholes [18].

Current characterization techniques used for the nanohole film quality-control are restricted to the inspection of very small areas, usually much smaller than the sensing chip (few squared micrometers at most). The inspection of larger areas is not easily affordable, because it implies high costs and long operation times. For instance, scanning electron microscopy (SEM) [21] and atomic force microscopy (AFM) [19, 22] have been used to perform geometrical and structural inspections of EOT based nanohole films. Also, near field scanning optical microscopy (NSOM) [2, 18] has been used to study the spatial distribution of the intensity of the light transmitted through this sort of films. Furthermore, although they give a high-resolution image of the nanostructured surface, they do not provide information about the plasmonic response of such surfaces. A very detailed information of the film nanostructures does not bring to a direct deduction of the spatial distribution of their plasmonic responses because the EOT phenomenon is collective in nature.

The spectral analysis of the light transmitted through the nanohole arrays in metallic films is a simple and fast way to characterize their EOT properties, preserving their collective nature; specifically, through the identification of the wavelengths at which the maximum light transmission occurs.

The nanohole arrays characterization done by independent measurements of the transmission spectra in small regions along the whole metallic film is ideal for the identification of inhomogeneities within a sensing chip.

Unlike the first experimental report on EOT phenomenon measurements, based on the use of a spectrophotometer for the measurement of an average spectrum resulting from the whole contribution of the sensing surface [1], the characterization done by means of the transmission spectra of small regions independently illuminated, enables to determine the spatial homogeneity of the optical response, after scanning the whole sensing surface. This is important since most of the EOT based devices use illumination spots smaller than the sensing area. It means that some kind of scanning microspectrographic device could be ideal for the characterization of the sensing chips.

The development of microspectrographic devices dates from 1940 [23], but most of their use has been restricted to the study of organic samples. The development of the first spectrographic microscopes were reported between 1940 and 1953 [23-26]. Their spatial resolution, in the visible spectral region, was close to 0.1 mm² [26], which is not enough to be used nowadays for the characterization of nanohole arrays. A similar device was reported in 1964 [27]; even if it was not able to scan surfaces, it was optimized to deal with inorganic samples under high pressure conditions and to work in both, the visible and the near-infrared spectral regions, but with spatial resolution as low as 1 mm². Later, in 1976, an electrodynamic condenser scanning method was reported to perform faster and more accurate microspectrographic scannings based on the displacement of the condenser lens over a fixed surface, but it was exclusively devoted to study subcellular organelles [28]. In 2007, the development of a spectrographic microscope for simultaneous absorption or emission *in-vivo* measurements on different subcellular compartments was reported too [29]. This device has a higher resolution and stability than all the previous ones, but it is not able to scan the sample. It uses two bundle light-guides for sample study, but its design and optimization are focused just for the optical absorption of biological samples.

There are just few reports on the study of nanohole patterned surfaces by means of microspectrographic methods. The first one dates from 1999 [20]. That work was focused onto the influence of adding a dimple lattice within a lattice of nanoholes in areas as small as few square micrometers. The work concluded that the total spectral transmission intensity is a function of the relative position between the nanoholes and the dimples, but that work did not consider scanning larger nanopatterned areas neither to use the spectral shifts for the characterization of the surface as we propose in this paper.

Later, in 2007, it was also reported the periodic modulation of transmitted intensity through subwavelength hole arrays surrounded by surface resonant cavities [30]. In that work the EOT signal intensity from the whole arrays, was measured using a microscope spectrophotometer, but it was not done a scanning of the arrays by means of the illumination of small regions, neither it was studied the maximum light transmission wavelengths.

A more recent work reports an innovative label-free optofluidic biosensor for single-cell cytokine secretion analysis based on nanohole arrays [31]. In that work, the characterization of their 4-inch diameter wafers patterned with uniform nanohole arrays was done by SEM. Wafers were further diced into 1x1 cm² sensor chips for cell analysis through optical detection, based on a spectroscopic imaging method which converts each imaged point into an effective sensing element. Thanks to the system we are proposing in this paper not only the nanohole patterned wafer characterization could be done faster and easier, but also the correct response of every effective sensing element could be ensured.

Bearing in mind the previous considerations, the development of an optical inspection tool which bridges the lab-industry gap is highly desirable.

Here, we describe the adaptation of an optical microscope to perform scanning spectrographic microscopy in large areas of the sensing chip surfaces. We suggest its use to check the homogeneity of the spectral response along all the nanohole patterned surfaces. It would require less time and costs than those required by other approaches.

The proposed method has an additional feature, it allows to work with nanohole patterned surfaces already embedded in microfluidic cartridges, because its operation is based on the optical far-field regime. Thus, the sensing chips can be calibrated *in-situ,* avoiding any damage introduced during a chip encapsulation process and enabling their later use for the detection of the specific position of biochemical events or for the detection of biological specimens by means of transmitted light spectral shifts. Furthermore, it can be used to compare signals coming from different channels in a multiplexed system and consequently, to identify the detected specimen.

We consider that this method could speed up the quality-control of the nanohole patterned surfaces, bridging the gap between lab and technological devices. In particular, the mass production of these nanopatterned surfaces and their use in biosensing devices could be benefited from our proposed inspection technique.

This manuscript is organized as follows. First, a Materials and Experimental section where it is described: (A) the features of the nanohole patterned surfaces used in this work, (B) the optical set-up, (C) the set-up illumination features, and (D) the set-up reliability. Then, a Results section where it is discussed: (A) the reproducibility of the set-up measurements and (B) the set-up performance in comparison to SEM. Finally, the main conclusions are reported.

## 2. Materials and experimental

### 2.1 Metallic nanohole patterned surfaces

The performance of our device was tested using nanopatterned surfaces made by Nanoimprint Lithography, according to a previously reported process [32]. Briefly, a silicon stamp fabricated by electron beam lithography and reactive ion etching was replicated into a thermal imprint resist (mr-I-7030, purchased to Microresist Technology GmbH), coating a 4" glass wafer. Prior to an imprinting step, the master was fluorosilane coated from vapor phase to facilitate the demolding. Following the imprinting step, an oxygen plasma etching was used for residual layer etching and pattern transfer to a glass substrate. Afterwards, a metallic Ti /Au layer was deposited by e-beam evaporation and the resist lift-off process was carried out in an ultrasonic hot acetone bath to obtain the nanoholes array structures. The described manufacturing protocol can be extended to 4" diameter surfaces, provided the stamp contains this large area nanopatterned array or a Step and Flash imprint tool [33] with smaller stamp surfaces. Larger areas can be also nanopatterned subject to the availability of suitable tool sizes for plasma etching and vapor deposition.

The nanopatterned surfaces are 50 nm thick gold films (99.999%) deposited on a glass slide coated with a 3 nm Ti film. Ti works as an adhesive between gold and glass. The nanostructures are square periodic arrays of circular nanoholes. The period of the square array is 500 nm and the average diameter of the nanoholes is 230 nm which are standard parameters to give a sensitivity of the order of 400 nm/RIU for a wavelength of 750nm for the maximum transmitted intensity at the plasmon resonance [34].

A simple microfluidic system was used in order to inspect the nanohole patterned surfaces while immersed in a liquid medium. It consists of a Teflon based housing with inlet and outlet tubes, and microfluidic channels made in a sandwich configuration with two transparent biocompatible polymer films: polycarbonate (PC) and double-sided pressure sensitive adhesive tape (PSA). The microfluidic channels were fabricated using a Silhouette Cameo cutting plotter. Figure 1 shows such microfluidic system.

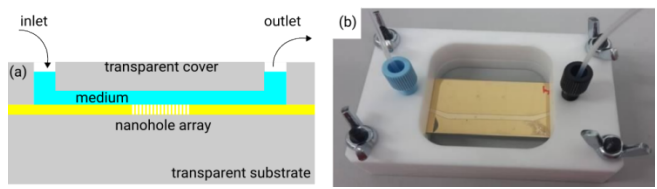

Fig. 1. (a) Schematics of the microchannel used to control the medium in contact with the nanohole arrays. (b) Picture of the actual microfluidic cell.

## 2.2 Scanning spectrographic microscopy set-up

The set-up for the inspection of the nanopatterned surfaces is a scanning spectrographic microscope. This device is able to measure the transmission spectrum of the nanopatterned surfaces in regions as small as few squared µm. The diameter enclosing such regions is going to be denoted by $D$ from now on. Besides, the device is also able to scan the nanopatterned surface, measuring the spectra of regions of diameter $D$, at a rate of 16 µm²/second. The results can be visualized as maps of the nanopatterned surface, where $D \times D$ µm² pixels are associated to the maximum spectral wavelength in the restricted regions where the spectra were measured.

The experimental set-up is depicted in figure 2. This is a modified version of an optical upright Nikon-Eclipse Ni transmission microscope, with the following features indexed as they appear in figure 2: a) the light source is a 100 W halogen lamp; b) the minimum field stop diameter is 1.5 mm; c) it has a geometric tailorable pinhole; d) the condenser is an achromatic and aplanatic lens with 0.5 numerical aperture and object distance equal to 1.6 mm; e) the sample-holder is motorized for displacements as small as 100nm along X and Y directions, and it is governed by means of a Prior-Scan III controller; f) the nanohole patterned sample; g) the objective lens is a 20x bright field lens; h) the collecting lens is a 4.5 mm focal distance lens; i) it has a 200 µm core Ocean Optics optical fiber, optimized for its use in the UV-visible spectral region; and j) an Andor Shamrock spectrograph operating with a 300 µm entrance slit and coupled to an Andor IDUS CCD camera with an integration time of 0.1 seconds. To increase the signal-noise ratio, the spectra are the result of the accumulation of 60 consecutive spectral measurements.

In this set-up, the pupil from the microscope built-in field stop is modified by means of the incorporation of a tailored pinhole (denoted as *field pinhole* from now on), placed at a distance $\Delta z$ above the field stop. The condenser lens is shifted to a new position placed at a distance $\Delta z'$ above its built-in position in such a way that the new distances from the nanohole patterned surface to the condensing lens and to the *field pinhole* produces a 0.1 lateral magnification. Therefore, if the light passes through a circular *field pinhole* of diameter *D'*, it is focused on the nanopatterned surface in the form of a *D=D'*/10 diameter circular spot.

The light transmitted through the nanopatterned surface is collected by the objective lens and, by means of the collecting lens, it is focused on the entrance of the optical fiber to be transmitted to the spectrograph where its spectrum is acquired.

Figure 2 summarizes all the optical set-up features. Its inset also remarks the modifications made to the original design of the commercial upright microscope. The introduction of a *field pinhole* in the optical design opens the possibility to control the shape and size of the illumination spot. The restriction in the area of illumination allows to inspect locally the nanopatterned surface with an increased sensitivity, because the transmission spectra are sensitive to the local characteristics (optical, geometrical, etc) of the illuminated region.

For a biosensing purpose, changes in the local effective refractive index in the illuminated volume close to the nanopatterned surface, depend on the optical properties and volume of any "imperfection" (biological or simply due to the manufacturing process) with respect to an ideal sensing chip. Thus, a reduced illuminated area increases the fill factor due to that "imperfection", modifying the effective refractive index of the volume close to the sensing surface.

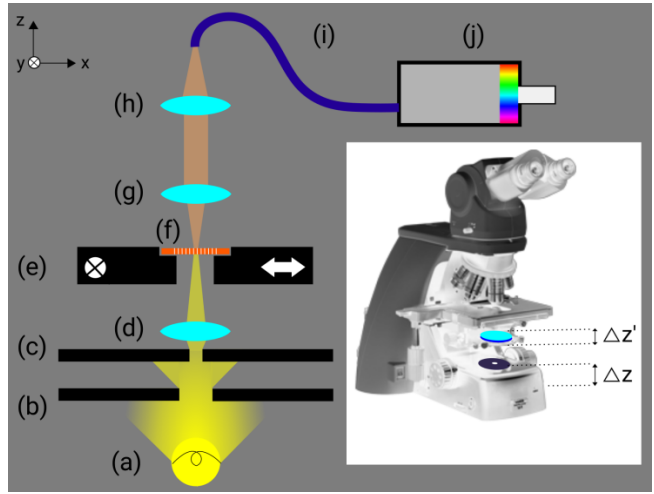

Fig. 2. Set-up schematics where (a) is the light source, (b) is the field stop, (c) is the *field pinhole*, (d) is the condensing lens, (e) is the motorized sample-holder, (f) is an EOT film, (g) is the objective lens, (h) is the collecting lens, (i) is the optical fiber and (j) is a spectrograph. The inset shows the pinhole and condensing lens positions in the microscope.

## 2.3 Light spot

This set-up is intended to measure the transmission spectra of the light passing through circular regions of diameter *D* of the nanopatterned sensing surface. In order to check the performance of our method, the value of *D* was fixed to 20µm. This was chosen looking at the sensor capability for sensing micron-sized biological objects randomly distributed on its surface, like for instance some of those contained in a liquid biopsy (leukocytes, possible cancer cells, etc). As it is shown in figure 3 for $D$=20μm, the intensity of the incident light on the nanopatterned surface is spatially distributed following a circular top-hat profile. Most of the light intensity is distributed in a 20 μm diameter circular light spot. The light passing outside such spot comes from diffraction effects at the pinhole aperture borders, but its intensity is negligible with respect to the intensity of the light inside the circular spot.

It means that each measured transmission spectra corresponds to circular regions of $D$=20μm in diameter. As mentioned above, the optical design is flexible enough to be used with other shapes and sizes of the illumination spot. This requires just to change the *field pinhole*.

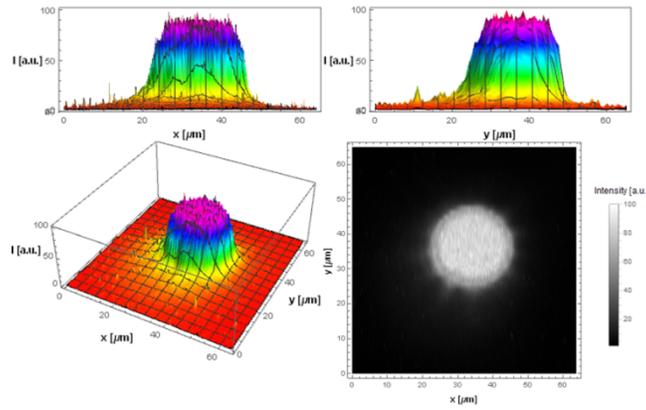

Fig. 3. Normalized spatial intensity profile of the illumination spot at the EOT surface. Top left: intensity vs x-coordinate plot. Top right: intensity vs y-coordinate plot. Bottom left: 3D representation of the spatial intensity profile. Bottom right: density plot of the intensity as function of x and y coordinates.

## *2.4 Set-up reliability*

We determined the set-up temporal stability in order to know the maximum error associated to its measurements. The temporal stability was determined by means of the measurement and analysis of the spectra of the light passing always through the same region of the sensing chip, which was continuously illuminated along 300 minutes. The figure 4 shows a superposition of all the acquired spectra as well as the evolution of the maximum spectral wavelength as function of time. The spectra were smoothed by means of a bilateral filter in order to reduce noise and to extract their maximum spectral wavelength. The considered maximum spectral wavelength is the one for the (1,0) medium-gold interface, at around 750 nm when the medium is water.

The maximum spectral wavelength range along 300 minutes was 0.545 nm, when the sensing chip is in water. And its corresponding standard deviation was equal to 0.151 nm. These values set the maximum error in the data coming from the set-up.

It is important to note that these results were acquired along a period of time larger than the typical one required for a usual scanning of a 260 x 260 μm$^2$ square nanopatterned surface.

The origin of the data fluctuations along the operation of the set-up may reside in a mixture of the following factors inherent to the set-up design: component mechanical stability, spectral and intensity stability in the illuminating source of light, and temporal stability in the spectrograph CCD.

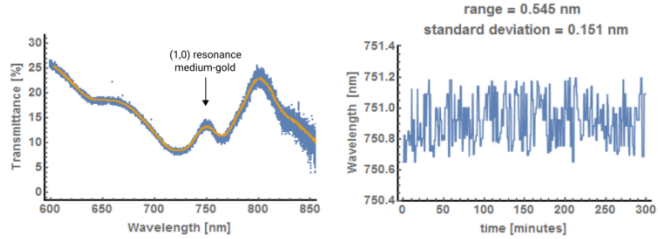

Fig. 4. Left: superposition of 300 transmittance vs. wavelength spectra taken from the same nanopatterned surface region during 300 minutes of continuous illumination. The blue points are experimental data, while the orange lines are the smoothed data. Right: Plot of maximum spectral wavelength vs illumination time; the blue line joins the experimental data.

## 3. Results and analysis

Statistical analysis of the measurements on nanohole patterned surfaces were carried out to know the optical set-up performance. Besides, the application of the optical set-up to find imperfections and impurities on sensing chips was compared with the capabilities of optical microscopy and SEM for the same purpose.

### 3.1 Optical scanning of nanohole patterned surfaces

Ten independent microspectrographic scannings were carried out on a sensing chip immersed in water, by means of the microfluidic system described in section 2. The sensing chip looked mostly homogeneous under the optical microscope and almost free of imperfections after a SEM inspection, as it is shown in figure 5.

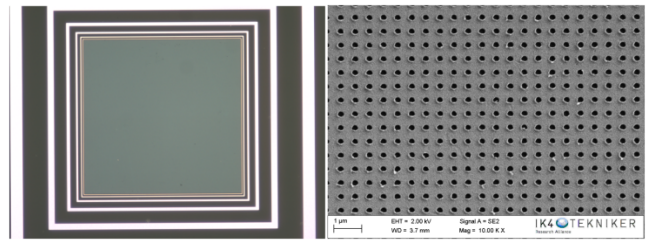

Fig. 5. Left: 10x optical microscope image of the nanopatterned surface. Right: SEM image of the nanopatterned surface.

The whole sensing chip was scanned in steps of $D$=20μm circular illumination regions. From all the measured spectra, the local maximum spectral wavelengths were determined for each position. A map of the mean maximum spectral wavelengths, along the chip, provides an image of the nanopatterned surface homogeneity. This map and its corresponding values are shown in figures 6a and 6b, respectively.

Just for the sake of clarity, the maps in figure 6 and subsequent figures are plotted with 20 μm x 20 μm squares instead of 20 μm diameter circles. However, the set-up allows to interchange easily the *field pinhole*, if either other shape or size of the illuminating spot were more suitable.

The map in figure 6a shows tiny differences in the optical response of different regions of the nanopatterned surface. The different resonance wavelengths correspond to slight inhomogeneities in the chip surface, either from structural defects or from impurities on its surface, which are not possible to detect directly by optical microscopy inspection. The histogram in figure 6b helps to understand how much slight the differences in the optical response of the chip sensing surface are.

On the other hand, a map of the standard deviation associated to each one of the values shown in the spectroscopic map of figure 6a gives the uncertainty associated to each one of the values. Such auxiliary map and its values indicate how much reproducible the measurements done by the set-up are.

The auxiliary map and its values are shown in figures 6c and 6d. It is noteworthy the high reproducibility of the measurements. The maximum standard deviation is just 0.45 nm, which is not a larger value than the maximum range associated to the set-up temporal stability.

Analogously, the optical sensitivity along the whole sensing chip was determined too. Scannings over the whole sensing chip were carried out with the chip immersed in several media, each one with a different refractive index. Five independent scannings were done for each one of the media. These were sucrose/water solutions at: 0, 20, 40 and 60 wt%, whose respective refractive indexes are 1.3333, 1.3620, 1.3907 and 1.4178. The optical sensitivity was calculated in nm/RIU units by least mean squares fittings of the maximum spectral wavelengths for each scanning position as function of the medium refractive index.

The figures 6e and 6f show the optical sensitivity values along the sensing chip while the figures 6g and 6h give an idea of their uncertainties in terms of their corresponding standard deviations.

Even if the nanohole patterned surface looks homogeneous under the optical microscope, the optical set-up is able to detect different optical sensitivities along the EOT surface. As much as we know this is the first time that a scanning map of optical sensitivities is reported for an EOT surface; it is usually reported the average optical sensitivity of the whole nanopatterned surface.

Maps of the optical responses of EOT surfaces are of paramount importance to perform the quality-control of their fabrication and to have a background reference of those surfaces when used as sensors of discrete moieties inhomogeneously distributed inside a test fluid medium. It is possible to see, from figures 6g and 6h, that most of the standard deviations have similar values, it is due to the optical set-up measurements reproducibility.

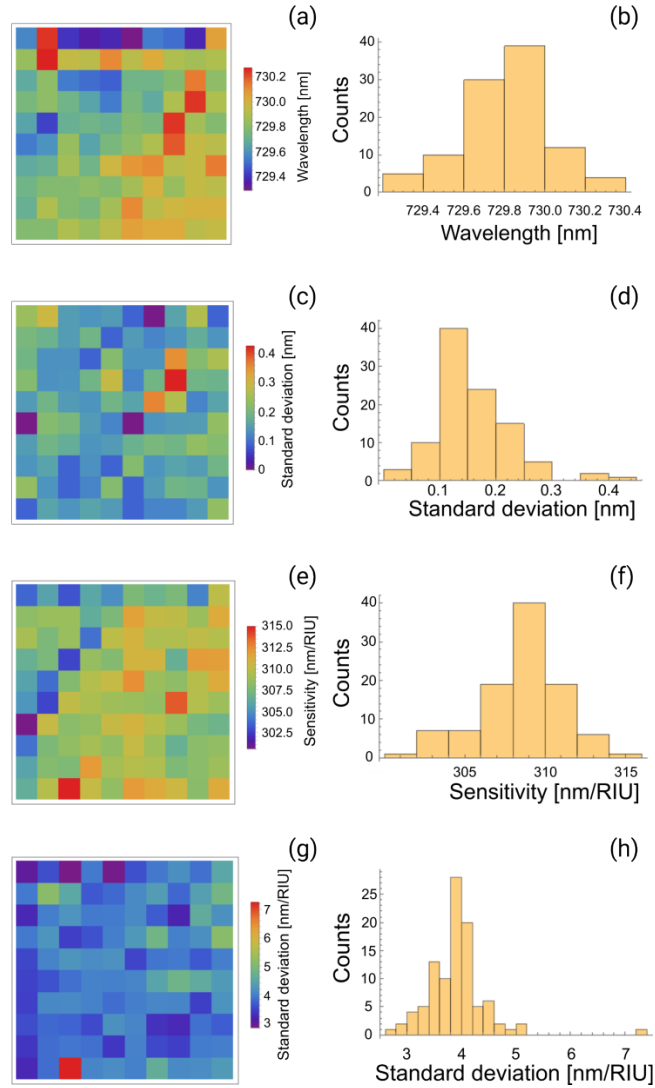

Fig. 6. (a) Map of the sensing chip with the corresponding mean maximum spectral wavelengths and (b) histogram of the its values. (c) Auxiliary map of the sensing chip with the standard deviations associated to the mean maximum spectral wavelengths, and (d) histogram of its values. (e) Map of the sensing chip with the corresponding mean optical sensitivity, and (f) histogram of its values. (g) Auxiliary map of the sensing chip with the standard deviations associated to the mean optical sensitivity, and (h) histogram of its values (g).

### *3.2 Scanning spectrographic microscopy as a bridge between the optical microscopy and the scanning electron microscopy*

Due to the intrinsic limitations in the fabrication of EOT surfaces by any nanofabrication technique, these are not always mostly homogeneous. Sometimes, the imperfections are clearly visible under the optical microscope, but there are many situations where imperfections are only detected through SEM. However, in both cases, just the detection of the imperfections is not enough to know their effect on the EOT phenomenon. Due to the collective nature of this phenomenon, sometimes not all the imperfections destroy the effect locally. Our set-up is able to detect imperfections as well as to determine their effect on the plasmonic performance in regions of the nanopatterned surface where the illumination spot is restricted.

The set-up is able to detect large inhomogeneities in the sensing chips, as much as an optical microscope may do, but it also points out such of them which really destroy the EOT phenomenon. This feature highlights the set-up as an attractive alternative for the quality control of the sensing chips.

For example, in an analogous way to what was described in the previous section, several spectrographic scannings were carried out on a nanohole patterned surface with randomly distributed imperfections. First, the sensing chip was scanned in water to determine its homogeneity, then the optical sensitivies along the sensing chip was determined.

In the first case, after ten independent scannings, the optical set-up was able to detect the large surface inhomogeneities through maps based on the mean maximum spectral wavelength values. Their corresponding standard deviations are directly related to the reproducibility of the measurements. In figure 7 these maps and their values clearly distinguish between regions with and without imperfections.

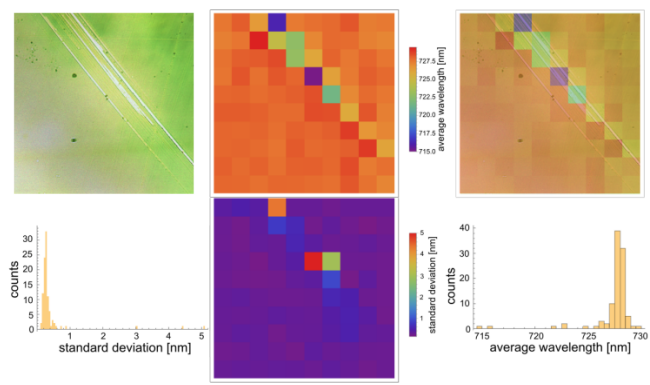

Fig. 7. Top left: 10x optical picture of the nanohole patterned surface in water. Top middle: map of the average of the maximum spectral wavelength. Top right: superposition of the top left and top middle images. Down left: histogram of the standard deviation values. Down middle: map of the standard deviations corresponding to the top middle map. Down right: histogram of the average of the maximum spectral wavelength values.

In the second case, the determination of the optical sensitivities was done following the same procedure described above for the mostly homogeneous surface. The figure 8 shows the maps and the values corresponding to the inhomogeneous nanopatterned surface. Again, the maps locate the position of the large surface imperfections in a reproducible way.

In figures 7 and 8, it is shown the correspondence between the spatial positions of the large imperfections and their effect on the optical response, which is clearly different to the optical response of the homogeneous regions. In both figures it is also evident that not all the imperfections modify the optical response in the same way. This is due to the collective nature of the EOT phenomenon; actually, the particular results shown in the figures correspond to the illumination of around $10^3$ nanoholes for each pixel in the map.

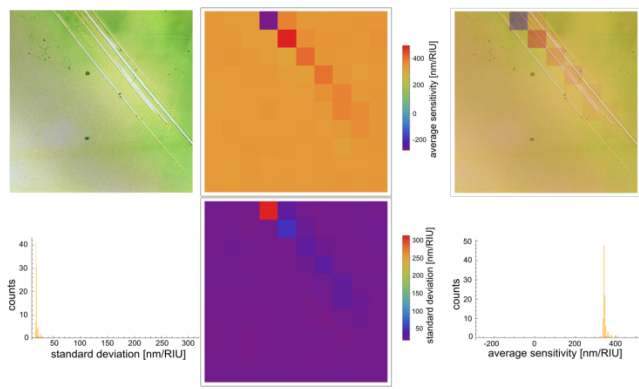

Fig. 8. Top left: 10x optical picture of the nanohole patterned surface in water. Top middle: map of the optical sensitivity. Top right: superposition of the top left and top middle images. Down left: histogram of the standard deviation values. Down middle: map of the standard deviations corresponding to the top middle map. Down right: histogram of the optical sensitivity values.

The inspection of nanohole patterned surfaces with imperfections so small that they are not resolved by optical microscopy is usually carried out by SEM. The optical microscopy has not enough spatial resolution to resolve nanometric motives because diffraction limitations, but the SEM high spatial resolution is ideal to perform detailed inspections of the quality and homogeneity of the nanoholes distributed in small areas, however, its associated high costs, large operation times and limitations to scan large surfaces makes it unavailable for practical purposes. The scanning spectrographic microscopy fills the gap between the optical microscopy and SEM, it is useful for the inspection of homogeneity in large EOT surfaces, even if they are already embedded in transparent microfluidic cartridges.

The scanning spectrographic microscopy set-up allows to control the quality of the nanostructures through the acquisition and analysis of the transmission spectra coming from regions restricted to 20 μm in diameter. After a scanning along the whole surface, the homogeneity of the optical response due to the nanoholes may be determined through the analysis of the collected spectra. With this technique, the operation time and costs associated to the inspection of surfaces as large as few $cm^2$ in area are much lower than those associated to the use of SEM. In comparison to the optical microscopy, this technique provides information about the nanopatterned surfaces beyond the diffraction limited optical microscopy resolution, because the transmission spectra measured by this technique are related to the quality of the nanoholes, whose sizes are shorter than the optical wavelength.

Figure 9 shows how the scanning spectrographic microscopy is able to locate the position of defects and pollutants in an EOT surface in coincidence with the information from both optical microscopy and SEM.

In Fig. 9, the pollutants in the nanopatterned surface are surrounded by a red square. The set-up detects the presence of pollutants on the surface, because they change the effective refractive index close to it, shifting the maximum spectral wavelength.

In a similar way, the fabrication defects are detected too. In figure 9, the surface fabrication defects are surrounded by blue and green squares. In this case the shift of the maximum spectral wavelength is less evident, because the illumination spot size is considerably larger than the nanoholes sizes; in such situation, the collective nature of the EOT phenomenon damps the effect of the defects in some individual nanoholes. The results obtained from the set-up preserves the information coming from the collective nanoholes optical response, this feature is completely unreachable by SEM and optical microscopy techniques.

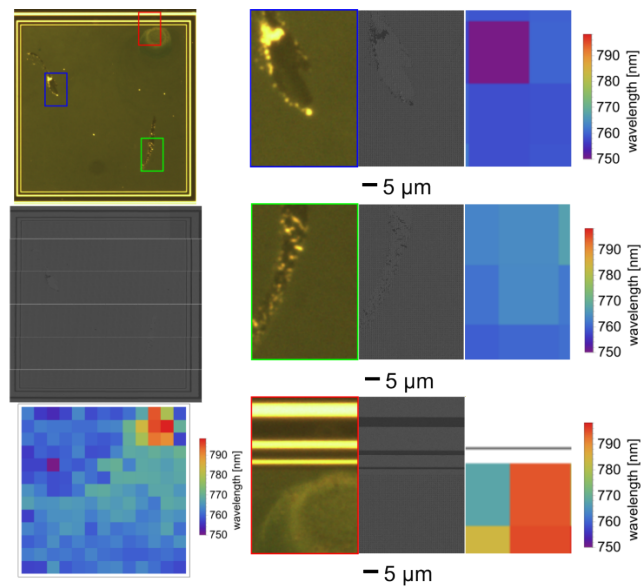

Fig. 9. Pictures of a 260 μm x 260 μm sensing chip in water. Top left: 10x optical microscopy picture of the film. It is clear the presence of few imperfections in the nanopatterned surface, some of them are surrounded by squares. The blue and green squares are for imperfections coming from the fabrication process. The red square is for a pollutant deposited on the surface. Middle left: Superposition of the SEM pictures taken over the whole film. The colored squares are the same size than each individual SEM picture. Bottom left: Map of the measured maximum spectral wavelength along all the film. The side of each pixel is 20 μm. Right: Zoom of the regions surrounded by the colored squares comparing the information coming from each individual technique.

However, even if the set-up preserves the collective behavior of the optical response, the restricted size of the illuminated regions still ensures enough high sensitivities. This makes an important difference with respect to other methods reported for the EOT measurements.

Figure 10 shows the scanning spectrographic microscopy ability to locate the position of very small defects in EOT surfaces. In this case, the defects are not large enough to be detected by optical microscopy (figure 10a), but they can be observed by SEM and also by the scanning spectrographic microscopy, as can be seen in figures 10b, c and d.

In figure 10, the position of the largest defect detected by SEM has been clearly identified by the scanning spectrographic microscopy. Such identification comes from a spectral shift at that position. The spectral map shown in figure 10d, evidences the different optical response at the top central pixel of the EOT scanned surface, which corresponds to the same position detected by SEM and shown in figure 10b. The values in the spectral map are the mean values from five independent scannings.

Figure 10e shows the measurement error associated to each value reported in the spectral map. They are expressed as the standard deviation of the five independent scannings.

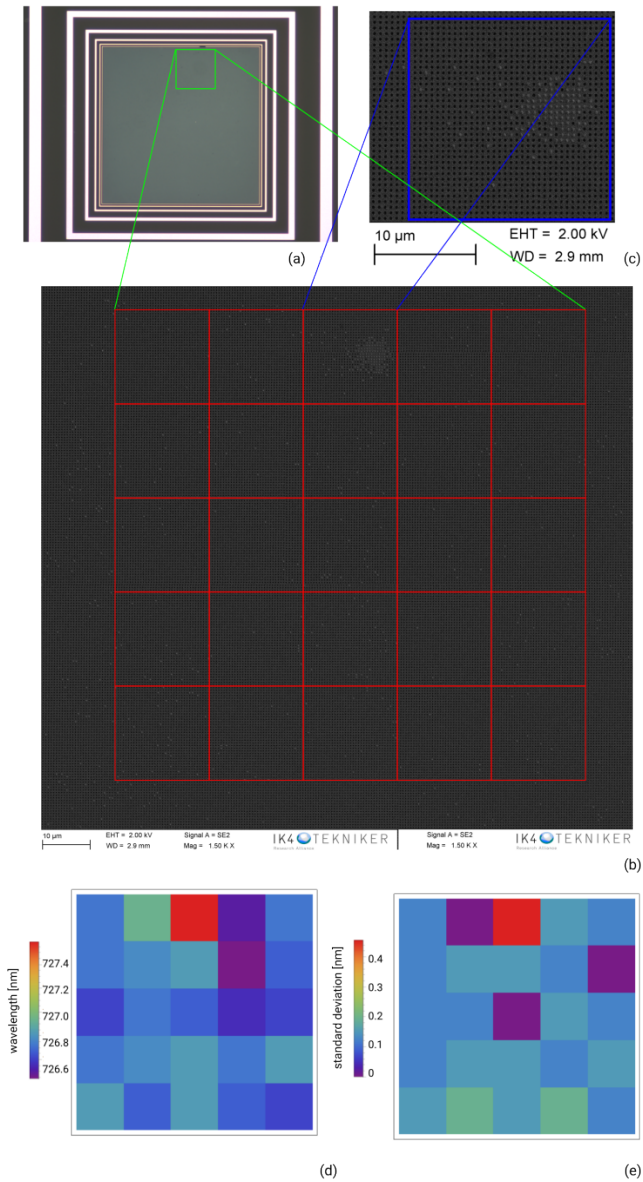

Fig. 10. Pictures of an almost homogeneous region of a 400 μm x 400 μm EOT film. (a) 10x optical microscopy picture of the film. The scanned region is surrounded by a 100 μm x 100 μm red box. (b)

SEM image of the largest defects in the scanned region. The square in blue corresponds to a 20 μm x 20 μm pixel in the spectral map shown in (c). Its position in the spectral map is in the middle of the first row. (c) SEM image of the scanned region. The squares in the red grid are for each one of the different positions of the 20 μm x 20 μm pixels in the spectral map. (d) Spectral map showing the mean resonance wavelengths along all the scanned region. The pixel with the largest resonance wavelength value corresponds to the SEM image in (b). (e) Standard deviations associated to values in the map.

The *chessboard-like* map in Fig. 10d (and also in some of the previous figures) is the main output of this inspection technique. It can be obtained in a fast and reliable way. Its size may be as large as the whole size of any nanopatterned array, and its pixels size may be tailored, just considering that their shortest size is diffraction-limited. The *chessboard-like* map points out the homogeneity of the manufactured nanopatterned arrays. It can give effective and fast information about defect areas so that if necessary, it may guide more detailed inspections through SEM in the position of the pixels signaling the defect area (for instance, the red square in Fig. 10d). The *chessboard-like* map actually is a *seal of identification* of the manufactured nanostructured films. Once it is measured it can be attached to the film and characterizes it for further uses. When the films are used as biosensors, it also maps their plasmonic performance and sets the background signal for detecting, for instance, micron-sized biological material (cells) distributed non-homogeneously on its surface [12, 31].

## 4. Conclusions

In this paper, we have introduced an optical technique for physical inspection of metallic thin films with nanohole arrays, intended to be implemented at industrial level. It is based on the detection of the wavelength spectral variations of the maxima positions of the light transmitted by micron-sized illuminated areas, taking advantage of the Extraordinary Optical Transmission phenomenon. As compared with usual techniques like SEM or other related electronic microscopy techniques, our proposed method is fast and accurate enough as to give a full surface homogeneity map of the nanopatterned surface in a short time. Electronic microscopy techniques can give geometry details of very small areas, but they are not able to show how these behave from a plasmonic point of view.

Our results have been validated with periodic nanohole patterned surfaces manufactured by NIL techniques. They show that the proposed technique is reproducible and reliable as to give a seal of identification of the manufactured surface (a *chessboard-like* map) which can be used as a background for its further application as a sensing device, especially when micron-sized biological material (i.e. cells) are distributed non-homogeneously on its surface.

This method is easily industrializable and can be extended to the inspection of either periodic or non-periodic nanostructured surfaces.

## 5. Funding and acknowledgments

*5.1 Funding*

Elkartek Program 2017 (µ4F, KK-2017/00089), United States International Technology Center-Atlantic (award # W911NF-17-2-0023)

*5.2 Acknowledgments*

Authors want to express their gratitude to SODERCAN (Sociedad para el Desarrollo de Cantabria) and Vicerrectorado de Investigación of the University of Cantabria. A.I.B. wants to thank the University of Cantabria for her FPU grant.

### References

1. T. W. Ebbesen, H. J. Lezec, H. F. Ghaemi, T. Thio, and P. A. Wolff, "Extraordinary optical transmission through sub-wavelength hole arrays", Nature **391**, 667-669 (1998).
2. H. F. Ghaemi, T. Thio, D. E. Grupp, T. W. Ebbesen, H. J. Lezec, "Surface plasmons enhance optical transmission through subwavelength holes", Phys. Rev. B **58** (11), 6779-6782 (1998).
3. Q. Li, Z. Yang, B. Ren, H. Xu, Z. Tian, "The relationship between extraordinary optical transmission and surface-enhanced raman scattering in subwavelength metallic nanohole arrays", J. Nanosci. Nanotechno. **10** (11), 7188-7191 (2010).
4. A. G. Brolo, "Plasmonics for future biosensors", Nat. Photonics **6**, 709-713 (2012).
5. A. A. Yanik, M. Huang, O. Kamohara, A. Artar, T. W. Geisbert, J. H. Connor, H. Altug, "An optofluidic nanoplasmonic biosensor for direct detection of live viruses from biological media", Nano Lett. **10** (12), 4962-4969 (2010).
6. H. Im, H. Shao, Y. I. Park, V. M. Peterson, C. M. Castro, R. Weissleder, H. Lee, "Label-free detection and molecular profiling of exosomes with a nano-plasmonic sensor", Nat. Biotechnol. **32**, 490-495 (2014).
7. M. Z. Mousavi, H. -Y. Chen, H. -S. Hou, C. -Y. -Y. Chang, S. Roffler, P. -K. Wei, J. -Y. Cheng, "Label-Free Detection of Rare Cell in Human Blood Using Gold Nano Slit Surface Plasmon Resonance", Biosensors **5** (1), 98-117 (2015).
8. M. Kalinich, D. A. Haber, "Cancer detection: Seeking signals in blood", Science **359** (6378), 866-867 (2018).
9. M. M. Ferreira, V. C. Ramani, S. S. Jeffrey, "Circulating tumor cells technologies", Mol. Oncol. **10** (3), 374-394 (2016).
10. J. W. Menezes, J. Ferreira, M. J. L. Santos, L. Cescato, and A. G. Brolo, "Large-Area Fabrication of Periodic Arrays of Nanoholes in Metal Films and Their Application in Biosensing and Plasmonic-Enhanced Photovoltaics", Adv. Funct. Mater. **20** (22), 3918-3924 (2010).
11. Y. Chuo, D. Hohertz, C. Landrock, B. Omrane, K. L. Kavanagh, and B. Kaminska, "Large-Area Low-Cost Flexible Plastic Nanohole Arrays for Integrated Bio-Chemical Sensing", IEEE Sens. J. **13** (10), 3982-3990 (2013).
12. A. I. Barreda, D. Otaduy, R. Martín-Rodríguez, S. Merino, J. L. Fernández-Luna, F. González and F. Moreno, "Electromagnetic behavior of dielectric objects on metallic periodically nanostructured substrates". Opt. Express **26** (9), 11222–11237 (2018).
13. M. Najiminaini, F. Vasefi, B. Kaminska, and J. J. Carson, "Experimental and numerical analysis on the optical resonance transmission properties of nano-hole arrays", Opt. Express **18** (21), 22255-22270 (2010).
14. D. Hohertz, S. F. Romanuik, B. L. Gray, and K. L. Kavanagh, "Recycling gold nanohole arrays", J. Vac. Sci. Technol. A **32** (3), 031403 (2014).
15. C. Genet and T. W. Ebbesen, "Light in tiny holes", Nature **445**, 39-46 (2007).
16. J. Beermann, T. Søndergaard, S. M. Novikov, S. I. Bozhevolnyi, E. Devaux, and T. W. Ebbesen, "Field enhancement and extraordinary optical transmission by tapered periodic slits in gold films", New J. Phys. **13**, 063029 (2011).
17. A. M. Mahros and M. M. Tharwat, "Investigating the Fabrication Imperfections of Plasmonic Nanohole Arrays and Its Effect on the Optical Transmission Spectra", J. Nanomater. **2015**, 178583 (2015).
18. T. Thio, H. F. Ghaemi, H. J. Lezec, P. A. Wolff, and T. W. Ebbesen, "Surface-plasmon-enhanced transmission through hole arrays in Cr films", JOSA B **16** (10), 1743-1748 (1999).
19. J. Zhang, M. Irannejad, M. Yavuz, and B. Cui, "Gold Nanohole Array with Sub-1 nm Roughness by Annealing for Sensitivity Enhancement of Extraordinary Optical Transmission Biosensor", Nanoscale Res. Lett. **10**, 238 (2015).
20. D. E. Grupp, H. J. Lezec, T. Thio, and T. W. Ebbesen, "Beyond the Bethe Limit: Tunable Enhanced Light Transmission Through a Single Sub-Wavelength Aperture", Adv. Mater. **11** (10), 860-862 (1999).
21. C. Escobedo, "On-chip nanohole array based sensing: a review", Lab. Chip **13**, 2445-2463 (2013).
22. T. I. Wong, S. Han, L. Wu, Y. Wang, J. Deng, C. Y. L. Tan, P. Bai, Y. C. Loke, X. D. Yang, M. S. Tse, S. H. Ng, and X. Zhou, "High throughput and high yield nanofabrication of precisely designed gold nanohole arrays for fluorescence enhanced detection of biomarkers", Lab. Chip **13**, 2405-2413 (2013).

23. T. Caspersson, "II.—Methods For The Determination of the Absorption Spectra of Cell Structures", J. Microsc. **60** (1-2), 8-25 (1940).
24. T. Caspersson, "A universal ultramicrospectrograph for the optical range", Exp. Cell Res. **1** (4), 595-598 (1950).
25. T. Caspersson, F. Jacobsson, and G. Lomakka, "An automatic scanning device for ultramicrospectrography", Exp. Cell Res. **2** (2), 301-303 (1951).
26. T. Caspersson, F. Jacobsson, G. Lomakka, G. Svensson, and R. Säfström, "A high resolution ultra-microspectrophotometer for large-scale biological work", Exp. Cell Res. **5** (2), 560-563 (1953).
27. H. C. Duecker and E. R. Lippincott, "Assembly and Performance of a Double-Beam Microscope Spectrophotometer from Commercial Instruments", Rev. Sci. Instrum. **35** (9), 1108-1112 (1964).
28. P. A. Benedetti, G. Bianchini, and G. Chiti, "Fast scanning microspectroscopy: an electrodynamic moving-condenser method", Appl. Opt. **15** (10), 2554-2558 (1976).
29. V. Evangelista, M. Evangelisti, L. Barsanti, A. M. Frassanito, V. Passarelli, and P. Gualtieri, "A polychromator-based microspectrophotometer", Int. J. Biol. Sci. **3** (4), 251-256 (2007).
30. N. C. Lindquist, A. Lesuffleur, S. -H. Oh, "Periodic modulation of extraordinary optical transmission through subwavelength hole arrays using surrounding Bragg mirrors", Phys. Rev. B **76**, 155109 (2007).
31. X. Li, M. Soler, C. Szydzik, K. Khoshmanesh, J. Schmidt, G. Coukos, A. Mitchell, H. Altug, "Label-free optofluidic nanobiosensor enables real-time analysis of single-cell cytokine secretion", Small **14** (26), 1800698 (2018).
32. J. M. Martínez-Perdiguero, A. Retolaza, D. Otaduy, A. Juarros, S. Merino, "Real-time label-free surface plasmon resonance biosensing with gold nanohole arrays fabricated by nanoimprint lithography", Sensors **13** (10), 13960-13968 (2013).
33. T. C. Bailey, S. C. Johnson, D. J. Resnick. S. V. Sreenivasan, J. G. Ekerdt, C. G. Wilson, "Step and Flash Imprint Lithography: An efficient nanoscale printing technnology", J. Photopolymer Sci. Tech. **15** (3), 481-486 (2002).
34. A.-P. Blanchard-Dionne and M.Meunier, "Sensing with periodic nanohole arrays", Adv. Opt. Photonics **9** (4)**,** 891-940 (2017).